# High-repetition-rate and high-photon-flux 70 eV high-harmonic source for coincidence ion imaging of gas-phase molecules

**Jan Rothhardt[1,2], Steffen Hädrich[1,2], Yariv Shamir[1], Maxim Tschnernajew[1,2], Robert Klas[1,2], Armin Hoffmann[1,+], Getnet K. Tadesse[1,2], Arno Klenke[1,2], Thomas Gottschall[1], Tino Eidam[1,+], Jens Limpert[1,2,3], Andreas Tünnermann[1,2,3], Rebecca Boll[5], Cedric Bomme[5], Hatem Dachraoui[5], Benjamin Erk[5], Michele Di Fraia[8], Daniel A. Horke[6], Thomas Kierspel[6], Terence Mullins[6], Andreas Przystawik[5], Evgeny Savelyev[5], Joss Wiese[6], Tim Laarmann[5,8], Jochen Küpper[6,7,8], Daniel Rolles[5,9]**

[1]*Institute of Applied Physics, Abbe Center of Photonics, Friedrich-Schiller-University Jena, Albert-Einstein-Straße 15, 07745 Jena, Germany*
[2]*Helmholtz Institute Jena, Fröbelstieg 3, 07743 Jena, Germany*
[3]*Fraunhofer Institute for Applied Optics and Precision Engineering, Albert-Einstein-Straße 7, 07745 Jena. Germany*
[5]*Deutsches Elektronen-Synchrotron DESY, Notkestrasse 85, 22607 Hamburg, Germany*
[6]*Center for Free-Electron-Laser Science (CFEL), DESY, Notkestrasse 85, 22607 Hamburg, Germany*
[7]*Department of Physics, University of Hamburg, Luruper Chaussee 149, 22761 Hamburg, Germany*
[8]*Hamburg Center for Ultrafast Imaging (CUI), University of Hamburg, Luruper Chaussee 149, 22761 Hamburg, Germany*
[9]*J. R. Macdonald Laboratory, Department of Physics, Kansas State University, Manhattan, Kansas 66506, USA*
[+] *present adress:Active Fiber Systems GmbH, Wildenbruchstraße 15, 07745 Jena, Germany*
[*]*j.rothhardt@gsi.de*

**Abstract:** Unraveling and controlling chemical dynamics requires techniques to image structural changes of molecules with femtosecond temporal and picometer spatial resolution. Ultrashort-pulse x-ray free-electron lasers have significantly advanced the field by enabling advanced pump-probe schemes. There is an increasing interest in using table-top photon sources enabled by high-harmonic generation of ultrashort-pulse lasers for such studies. We present a novel high-harmonic source driven by a 100 kHz fiber laser system, which delivers $10^{11}$ photons/s in a single 1.3 eV bandwidth harmonic at 68.6 eV. The combination of record-high photon flux and high repetition rate paves the way for time-resolved studies of the dissociation dynamics of inner-shell ionized molecules in a coincidence detection scheme. First coincidence measurements on $CH_3I$ are shown and it is outlined how the anticipated advancement of fiber laser technology and improved sample delivery will, in the next step, allow pump-probe studies of ultrafast molecular dynamics with table-top XUV-photon sources. These table-top sources can provide significantly higher repetition rates than the currently operating free-electron lasers and they offer very high temporal resolution due to the intrinsically small timing jitter between pump and probe pulses.

## 1. Introduction

The availability of ultrashort-pulse short-wavelength photon sources is an essential prerequisite to enable studies of matter on atomic (picometer) length and (femtosecond) time scales. As such, they constitute an indispensable tool for modern science, which is reflected by the growing number of free-electron lasers (FELs) that are operating or being constructed at the moment [1–6]. These sources allow to address many of the fundamental questions in areas as diverse as physics, chemistry, biology, material science and medicine [7,8]. Here, we will concentrate on ultrafast molecular dynamics imaging experiments [9–11] that aim, e.g., at tracing initial charge-transfer processes in space and time [11,12], which can be considered as a fundamental process in many photochemical reactions. Understanding these processes has the potential to control, manipulate, and steer the reaction in a desired channel using ultrafast light. A particularly powerful method for such studies is the coincident momentum imaging technique [13], which yields multi-dimensional data sets that can provide kinematically complete, channel-resolved, and highly differential measurements that contain a wealth of information, and which allow studies in the molecular reference frame [11]. However, it imposes the restriction of less than one ionization event per X-ray pulse in the probed sample, on average, to unambiguously correlate the ionization fragments stemming from the same molecule. Therefore, the acquisition of sufficient statistics is only feasible with high-repetition-rate pump-probe sources, ideally with a multi-kHz repetition rate, which is beyond the capabilities of the current FEL setups. Moreover, the x-ray pulses must be synchronized to optical pump pulses with femtosecond precision, which is hindered by the long beam paths through the FEL machines and the intrinsic arrival time jitter [15]. Nevertheless, pump-probe timing jitter at FELs has steadily been improved and sub 10 fs timing jitter has already been demonstrated by a measure-and-sort method recently [16]. Complementary table-top sources, realized *via* high harmonic generation (HHG) of ultrashort-pulse lasers, have attracted significant attention with a growing user community over the last years, since they are compact and highly accessible. The process of HHG itself has been known for several decades already, but the repetition rate of the sources has been mostly limited to the low kHz regime, which is dictated by the employed laser technology [17–21]. However, there is a strong demand for high repetition rate sources not only for coincidence experiments, but also for many applications, e.g. in surface science [22,23] or atomic [24] and molecular [25] physics. The development of new high repetition rate HHG sources has seen rapid progress enabled by recent advances in high average power femtosecond laser technology [26–32]. In that regard, fiber lasers in combination with nonlinear compression have played a pivotal role, since these enabled high average power ultrashort pulse laser sources [33,34]. Additionally, the investigation of phase-matching aspects during the frequency up-conversion for such rather low pulse energy lasers has finally led to efficient HHG using fiber lasers [35,36]. As a consequence, this approach produced a maximum photon flux of more than $10^{13}$ photons/s (100 µW of average power) in the XUV at 600 kHz repetition rate [37], or $>10^9$ photons/s in the soft X-ray region at 100 kHz [38]. The repetition rate of these sources can even be pushed up to 10 MHz, still with $10^{13}$ photons/s [39], rendering novel applications in solid state and molecular physics possible with fiber-laser-based HHG sources.

In this contribution, we present coincidence experiments for iodomethane ($CH_3I$) that were conducted with 68.6 eV photons from a laser-based 50–100 kHz repetition rate source. The XUV source was enabled by a state-of-the-art chirped-pulse-amplification fiber-laser system that incorporates coherent combination of two main amplifier channels to achieve 1 mJ, 300 fs pulses with 50–100 W of average power at a central wavelength of 1030 nm. Subsequent spectral broadening in an argon-filled hollow-core fiber followed by temporal compression *via* chirped mirrors supplied 0.6 mJ pulses with a duration of 35 fs. A part of this pulse energy (up to 100 µJ) can be used for optical pumping of the molecular sample ($CH_3I$) while the remainder is used for high-harmonic generation. Particular optimization of the photon flux of the $57^{th}$ harmonic yielded up to $6.8 \cdot 10^{10}$ photons/s at a repetition rate of 50 kHz. This is not only the highest achieved value in this spectral range, but it also allowed for the first successful demonstration of a coincidence measurement on inner-shell ionized gas-phase molecules performed with table-top sources. Previous coincidence experiments with HHG sources have been limited to valence ionization [40–45] and only a few, non-coincident atomic inner-shell ionization experiments have been performed with HHG sources to date [46–48].

Important aspects for future pump-probe experiments, such as finding the temporal overlap between the pump and the probe pulses, increasing the density of the molecular beam, and advances in laser technology to further increase the photon flux will be discussed, providing evidence that time-resolved coincidence experiments on inner-shell ionized gas-phase molecules with high-order harmonic sources are in reach. Since HHG sources can also provide attosecond pulses at high repetition rates [49], such experiments are expected to access molecular dynamics at shortest time scales in the future.

## 2. High-repetition-rate and high-photon-flux XUV source

The realization of a long-term-stable high-repetition-rate source of high harmonics inherently requires a suitable high average power laser system. Here, a fiber chirped pulse amplifier was used that was operated with two coherently combined main amplifier channels and that is described in section 2.1. The pulses emerging from this laser were subsequently compressed in time utilizing a gas-filled hollow-core fiber and chirped mirrors, as outlined in section 2.2. After the compression step, high harmonics were generated in an argon gas jet with the generation process being particularly optimized for a photon energy of 68.6 eV. The experimental characterization of the XUV source and details on selecting the harmonic and focusing it onto the molecular target are described in section 2.3.

### 2.1 Coherently combined femtosecond fiber chirped pulse amplifier

The fiber chirped pulse amplification (FCPA) system is sketched in Figure 1. Its design is similar to the laser system reported in [50], but incorporates only two instead of four main amplifier channels. The frontend is a home-built femtosecond fiber oscillator that operates at 19 MHz repetition rate and 1030 nm central wavelength with a spectral bandwidth of ~10 nm (FWHM). It is directly spliced to a first fiber-integrated pre-amplifier, which increases the average power to 130 mW. Subsequently, the pulses are stretched to ~1.5 ns in duration by means of a dielectric grating based stretcher and then sent into a spatial-light-modulator-based pulse shaper, which is used to compensate the residual linear and nonlinear spectral phase for almost perfect pulse compression. The signal after this device is coupled to another cascade of fiber-integrated pre-amplifiers that also include two acousto-optical modulators which are used to reduce the repetition rate to 1 MHz and then to the final value of 50 kHz or 100 kHz for the present experiment; this yields a pulse energy of 0.5 µJ. This amplifier chain is followed by a first large pitch fiber (LPF) with a mode field diameter of 55 µm, which delivers 20 µJ of pulse energy for seeding of the main amplifiers. As shown in Fig.1 this signal is equally split into two spatial replicas by a polarizing beam splitter cube and each of the beams is sent to a large pitch fiber amplifier with 80 µm mode-field diameter. Before entering the main amplifiers, the linearly polarized signals are circularly polarized with a quarter wave plate (QWP) to reduce the impact of nonlinearity [51]. After amplification and traversing another QWP each channel delivers ~750 µJ pulses with linear polarization, and recombination is done with a thin film polarizer (TFP).

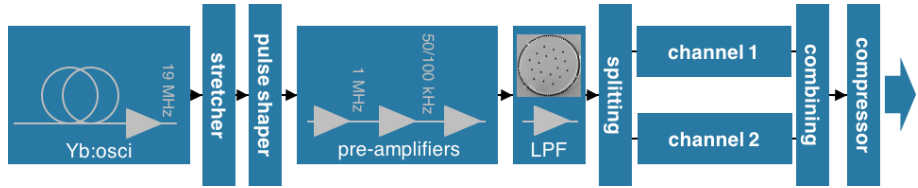

**Figure 1.** Experimental setup of the coherently combined fiber chirped pulse amplification system (CC-FCPA).

After recombination a small portion of the beam (<1%) is sent to a Hänsch-Couillaud detector, that analyzes the polarization state of the recombined beam and uses this to generate an error signal. This is fed-back to a piezo-driven mirror within the pre-amplifier section of the laser system that adjusts the temporal delay between the two channels [30,50]. Finally, the beam is compressed to 300 fs duration in a dielectric grating based compressor. The compressed pulse energy is 1 mJ, corresponding to an average power of 50 W (100 W) at 50 kHz (100 kHz) repetition rate. Due to the use of fiber amplifiers the spatial beam quality is excellent as verified by $M^2$<1.4 measurements in previous experiments [52,53].

*2.2 Nonlinear compression*

As outlined in section 2.1, the pulses emerging from the coherently combined-fiber chirped-pulse-amplification system (CC-FCPA) are 300 fs in duration, which is too long for efficient high harmonic generation [19]. For that reason a nonlinear compression (NC) setup has been implemented that closely follows the design presented in [34,37]. A schematic setup of this NC stage together with the pump-probe splitting is shown in Fig.2.

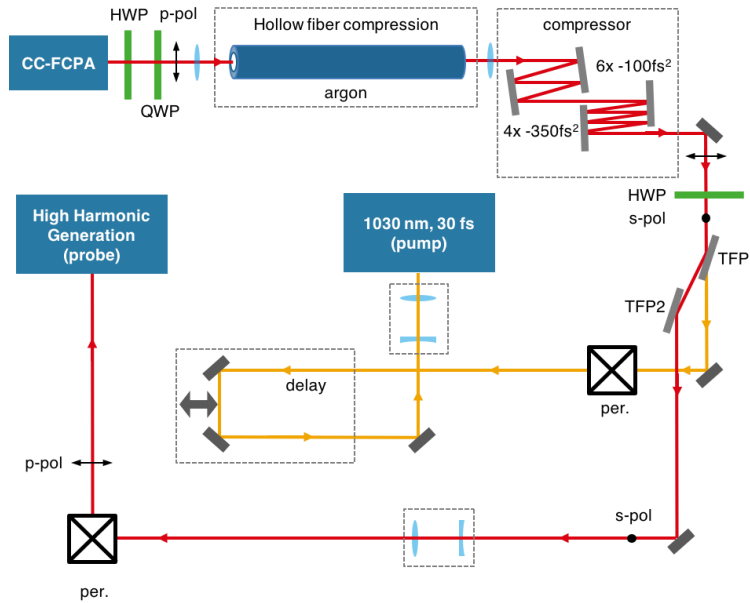

**Figure 2.** Experimental setup of the nonlinear compression stage and the pump and probe generation. As mentioned in the text p-polarization refers to parallel and s-polarization to perpendicular polarization with respect to the surface of the optical table. (HWP – half-wave plate, QWP – quarter-wave plate, TFP – thin film polarizers, per. - periscope)

The 1 mJ, 300fs pulses are coupled into an argon-filled (~4.5 bar) hollow-core fiber with an inner diameter of 250 µm and a length of 1 m. Propagation in this waveguide leads to self-phase modulation resulting in spectral broadening. The imposed chirp is removed by a chirped mirror compressor with a group delay dispersion (GDD) of -2000 fs$^2$ leading to 35 fs pulses. The pulse energy after the compressor is still as high as 0.6 mJ corresponding to an average power of 30 W (60 W) at 50 kHz (100 kHz) repetition rate. The polarization used for the NC stage is parallel to the optical table which is ensured by a combination of a quarter and half-wave plate located before the hollow-core fiber. In the following, we will refer to p-polarization (s-polarization) if the polarization is parallel (perpendicular) to the surface of the optical table. A measurement of the polarization revealed less than 1% of content in the perpendicular polarization state after the chirped mirror compressor. This linearly polarized light is sent through another half-wave plate (HWP) and then impinges on a chicane of two thin film polarizers (TFP), which are used to split part of the beam by rotating the polarization with the HWP (Fig. 2). The transmitted part with an energy of up to 100 µJ, which is p-polarized, is sent through a delay line and can then be used as a pump pulse, initiating electronic and nuclear dynamics in small molecular quantum systems. Although in the present experiment no time-dependent molecular-dynamics were observed, robust methodologies for finding temporal overlap were established as preparatory work, see supplementary online material for details. The main part of the beam with 0.5 mJ of pulse energy and s-polarization is reflected off the TFPs, traverses another telescope to enlarge the spatial beam diameter, is sent up on a periscope rotating back to p-polarization and then enters the high harmonic generation chamber. The latter will be described in the following section 2.3 and provides the XUV-probe beam at 68.6 eV used in present experiments.

*2.3 High photon flux source at 68.6 eV*

Using the 0.5 mJ, 35 fs pulses, which are available after splitting the pump beam (section 2.2), a high harmonic source has been realized that is particularly optimized for operation at 68.6 eV. This photon energy was chosen since it allows photo-ionizing the iodine *4d* inner-shell electron in the CH$_3$I molecule.

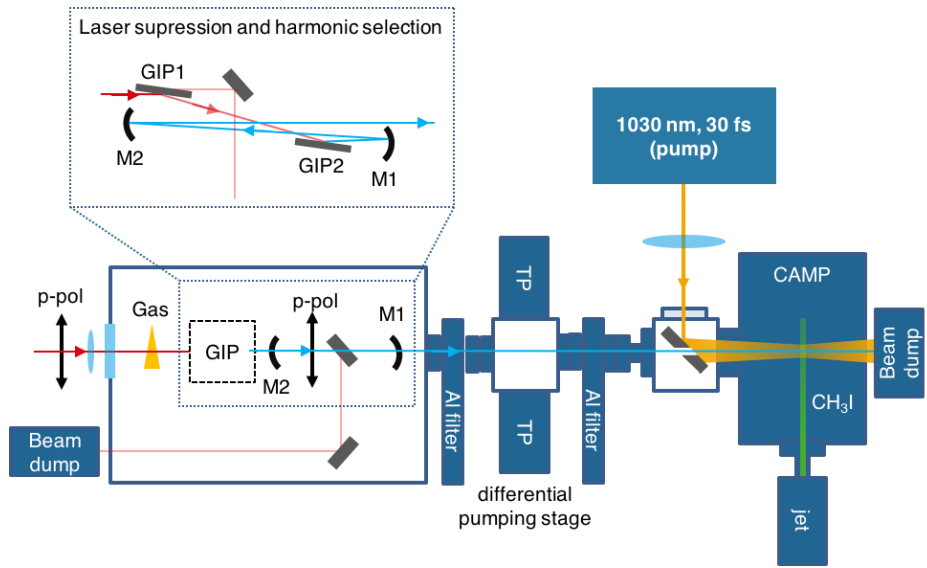

**Figure 3.** Top view on the experimental setup used for high harmonic generation. The inset shows a more detailed side-view of the separation of the harmonics from the fundamental driving laser, selection of a single harmonic and focusing of the harmonic into the experiment. (GIP – grazing incidence plates, M1/2 – XUV mirrors at 68.6 eV, TP – turbo pumps, CAMP – CFEL-ASG Multi Purpose end station [54]).

The experimental setup of the HHG source, including the selection of a single harmonic and the steering to the experimental chamber, is shown in Fig. 3. The pulses are sent into a vacuum chamber and are focused into an argon gas jet, which produced from a thin round orifice with 150 μm inner diameter. Using a $f = 300$ mm focusing lens leads to a focal spot size ($1/e^2$-intensity) of ~90 μm and an intensity of $>2\cdot10^{14}$ W/cm$^2$. The generated high harmonics and the laser co-propagate and impinge on a chicane of two grazing incidence plates (GIP) [55]. The latter are XUV-grade SiO$_2$ substrates that are anti-reflection coated for 930 nm – 1130 nm (s-polarization) used under 82° angle of incidence. As shown in the detailed view in Fig.3, the two GIPs are used such that the XUV and infrared are s-polarized with respect to the plane of incidence. Therefore, the XUV beam is changed in height, but maintains p-polarization with respect to the optical table and the detector normal (see section 3). The reflectance of the top layer of SiO$_2$ is calculated to be 63 % (40 % after two reflections) of the 68.6 eV photon pulses, whereas the reflection of the infrared laser is reduced to 10% by the AR coating. Consequently, the IR laser is suppressed to 1% after two reflections, which is important to prevent subsequent aluminum filters for complete reduction of NIR light from damage.

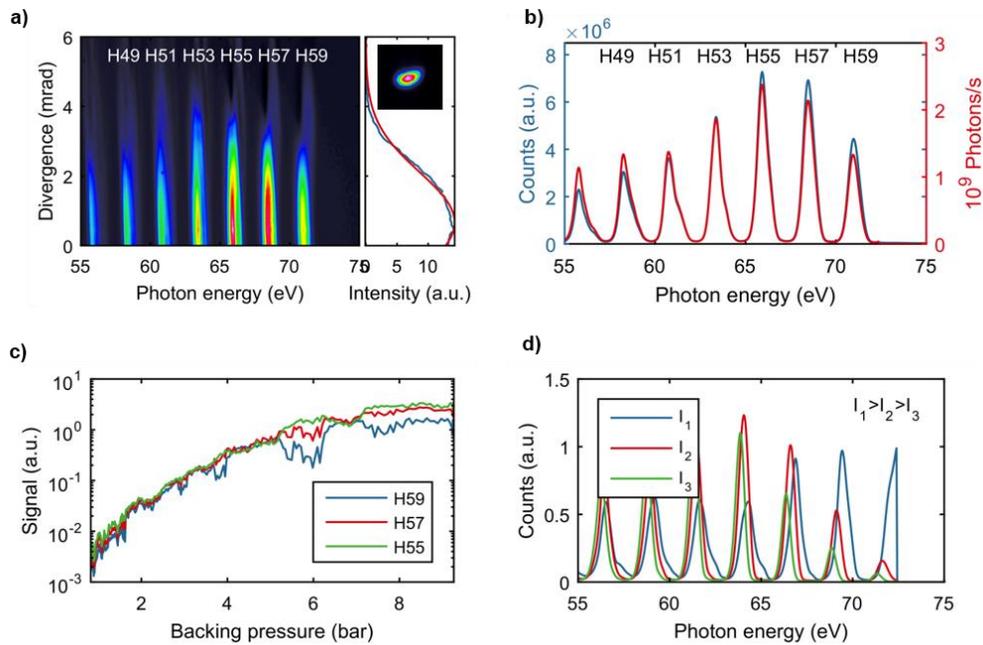

**Figure 4.** Characterization of the high harmonic source. **a)** Measured spatio-spectral distribution of the harmonics at the optimal generation conditions (7 bar backing pressure). The HHG signal has been analyzed with an imaging spectrometer (flat-field grating) that allows observation of the divergence for each individual harmonic. The right part of (a) shows the lineout (blue) and a Gaussian fit (red) of H57 at 68.6 eV. The inset shows the focus of the XUV beam at 68.6 eV. **b)** Integration along the spatial (divergence) axis leads to the harmonics spectrum (blue). When accounting for the detection efficiency and background absorption in the chamber a calibrated spectrum (red curve) can be obtained. **c)** Harmonic signal of H55, H57 and H59 vs. backing pressure. A saturation of the signals is reached at pressures above 7 bar. **d)** The harmonics can be slightly shifted by varying the intensity of the input laser. Three spectra for three different intensities are plotted.

In a first experimental campaign, a thorough characterization of the high harmonic generation source was performed. For that purpose, the XUV beam after the second reflection of the GIP was sent through two aluminum filters with a thickness of 200 nm each, which were used to further suppress the 1030 nm driving laser and to attenuate the XUV in order to prevent saturation of the following detector. A spatial-spectral analysis of the harmonics was done with a flat-field grating spectrometer with an attached XUV CCD camera. The experimental conditions, i.e., intensity, gas jet position relative to the focus, and the gas pressure were optimized so as to maximize the signal of H57 at ~68.6 eV. Figure 4 (a) shows the spatial-spectral distribution of the harmonic signal between 55 eV and 75 eV at optimum conditions (7 bar backing pressure). Under these conditions, spatially well-defined harmonics were generated with nearly Gaussian-like profiles (Fig. 4 (a) right side). Integration along the spatial direction yields the harmonic spectrum as shown by the blue curve in Fig. 4 (b), which can then be used to obtain the number of photons per second by accounting for the detection efficiency as described in [37].

Furthermore, absorption on the 90 cm beam path to the detector has been taken into account. The rather high background pressure of $4 \cdot 10^{-1}$ mbar in the chamber results in a transmission of only T=30% at 68.6 eV. Note that additional corrections due to clipping of the XUV beam at the detector, which is obvious in Fig. 4 (a), and at the diffraction grating are taken into account for estimation of the photon flux. The corrected spectrum is displayed in Fig. 4 (b) (red curve). The resulting photon flux for the highest harmonic orders is obtained by additional spectral integration across each individual harmonic line and summarized in Table 1.

**Table 1.** Number of photons per second generated in harmonic 47 to 59 measured by the grating spectrometer. The laser was operated at 50 kHz repetition rate.

| Harmonic | 47 | 49 | 51 | 53 | 55 | 57 | 59 |
|---|---|---|---|---|---|---|---|
| Photon energy | 56.6 | 59.0 | 61.4 | 63.8 | 66.2 | 68.6 | 71.0 |
| Photons/s ($10^{10}$) | 5.0 | 6.4 | 7.9 | 7.9 | 7.8 | 6.8 | 4.0 |

It should be noted that this is a rather conservative analysis and resulting in a minimum value for the achieved photon flux for the measured harmonics. It yields $6.8 \cdot 10^{10}$ photons/s for harmonic 57 at 68.6 eV, which constitutes at least an order of magnitude higher photon flux than previously reported in this spectral region [56,57]. However, the transmission of the Al filters, the diffraction efficiency of the grating and the detection efficiency of the XUV CCD tend to degrade with time due to surface contamination. Hence, the real photon flux might have been even higher than estimated above. As described below, an independent measurement of the photon flux utilizing photoionization within a neon flooded experimental chamber yields about $2 \cdot 10^{11}$ photons/s for H57. The values shown here are obtained at a repetition rate of 50 kHz, while part of the experiments (section 3) has also been conducted with a repetition rate of 100 kHz. Note that all other laser parameters such as pulse energy, pulse duration and focal spot diameter remain unchanged. Hence, the XUV photon flux is expected to increase linearly with the laser repetition rate [37] and $>10^{11}$ photons/s could be generated in a single harmonic at 68.6 eV with the presented system at 100 kHz.

Due to the use of grazing incidence reflections, the XUV beam still contains a broad spectrum of harmonics after the GIPs. Selection of a single harmonic (H57) is done by two XUV mirrors that additionally are used for collimation and re-focusing (Fig. 3). Both are designed for reflection of 68.6±0.5 eV (R=0.5), while the first mirror (f=300 mm) collimates the XUV beam and the second one (f=1200 mm) focuses the XUV pulses onto the molecular beam of $CH_3I$ (Fig. 3). The latter is located in the CAMP end station. The inset of Fig. 4 (a) shows a nice and clean XUV focal spot with a diameter of ~120 μm ($1/e^2$-intensity). Fine tuning of the photon energy of the harmonic so as to match the mirror reflection curve was possible by adapting the intensity of the driving laser. Three harmonic spectra for different driving laser intensities are shown in Fig.4 d). The HHG chamber was separated from CAMP via a differential pumping unit (Fig. 3) and a sealed aluminum filter with 200 nm thickness that was placed just in front of the differential pumping unit. This way, the pressure in CAMP stayed below $10^{-9}$ mbar, while the pressure was $4 \cdot 10^{-1}$ mbar in the HHG chamber when the gas jet was operated. As shown in Fig. 3, the XUV (probe) beam was focused into CAMP through a holey mirror while the infrared pump was reflected of the same device with a focusing lens placed outside of the chamber. The temporal delay was controlled by an additional delay line within the pump arm (see Fig. 2). Note that the employed 200 nm thick aluminum filter suppressed the average power of the HHG-driving IR beam to <0.3 mW for the coincidence experiment.

## 3. Coincidence detection of CH$_3$I ionization fragments

The coincidence experiment was performed in the CAMP instrument [54], a mobile multi-purpose experimental station that was previously used, e.g., for electron and ion imaging experiments at the LCLS and FLASH free-electron lasers [9,10,58,59] and for electron-ion and ion-ion coincidence experiments at the PETRA III synchrotron radiation source at DESY. The current setup consists of a doubly skimmed supersonic molecular beam (nozzle diameter: 30µm, skimmer diameters: 200 and 400 µm, distance nozzle to interaction region: roughly 70 cm) and a double-sided velocity map imaging (VMI) spectrometer equipped with two microchannel plate (MCP) detectors with position-sensitive delay-line anodes (Roentdek *DLD80* for the ions and *HEX80* for the electrons), as shown in Figure 5. The HHG and/or NIR beams intersected with the molecular beam inside the VMI spectrometer, where a constant, inhomogeneous electric field guided the electrons and ions created by the interaction of the photons with the target molecules onto the position-sensitive detectors. The MCP and position signals were recorded with a multi-hit time-to-digital converter (TDC), and the data was stored event by event in a list-mode format that allowed retrieving coincidences between the detected particles. From the time of flight and the hit positions of the ions, the three-dimensional ion momenta can be calculated, while the time of flight of the electrons is too short for a meaningful 3-D momentum reconstruction such that only 2-D projections of the electron momentum distributions were measured. For cylindrically symmetric electron distributions it is possible to retrieve the 3-D distributions through inversion algorithms [60,61].

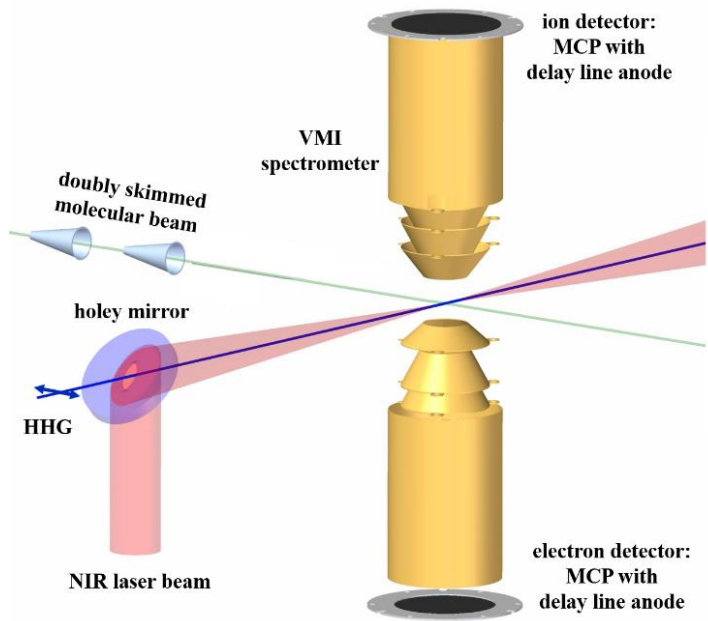

**Figure 5.** Schematics of the coincidence apparatus with a supersonic molecular beam and a double-sided velocity map imaging (VMI) spectrometer with MCP delay-line detectors for time- and position resolved coincident electron and ion detection.

In the experiment described here, we recorded electron-ion-ion coincidence data for the photoionization of $CH_3I$ molecules by photons from the HHG pulses. At a photon energy of 68.6 eV, the cross sections for both valence and inner-shell ionization at this photon energy are roughly equal [62]. Thus, the $CH_3I$ molecules can either be valence ionized, resulting predominantly in singly charged $CH_3I^+$ ions, or a I(4d) inner-shell electron can be removed, which leads to a doubly or triply charged final state after the inner-shell vacancy relaxes *via* Auger decay. The majority of these doubly or triply charged cations then break up into charged fragments as seen in the ion time-of-flight mass spectrum shown in Fig. 6 (a). A plot of the ion-ion coincidences between $CH_x^+$ (x = 0, 1, 2, 3) fragments and $I^+$ fragments is shown in Fig. 6 (b). Due to momentum conservation, these coincidence events, which resulted from a (quasi-) two-body fragmentation — the neutral or charged $H^+$ fragments that may also be emitted carry very little momentum due to their light mass —— can be clearly identified by diagonal lines in the photoion-photoion coincidence (PIPICO) spectrum. The corresponding detector images for $I^+$, $CH_x^+$, and $CH_3I^+$ are shown in Figure 7. Note that these results were obtained from a continuous 20-hour long measurement during which the laser system, the high harmonic generation, and the coincidence apparatus were operated constantly and without interruption. This reliability and long-term stability is an important prerequisite for the coincidence measurements presented here and for future experiments.

In order to demonstrate spatial and temporal overlap between the NIR-pump pulses and the XUV-probe pulses, we removed the Al filter in front of the differential pumping section such as the NIR drive-laser beam from the HHG chamber, which traveled collinear with the XUV pulses, could reach the interaction region inside the CAMP chamber. Using a small YAG crystal that could be driven into the interaction region on an xyz-manipulator and that could be viewed on a CCD camera through a long-distance microscope, we first spatially overlapped the two NIR beams. Then we used a fast photodiode that was also mounted on the xyz-manipulator to overlap the two laser pulses to within a few tens of picoseconds. Finally, we scanned the delay stage in the pump arm until we observed interference fringes on the YAG screen due to interferometric autocorrelation between the two NIR pulses, as shown by the short movie that we recorded with the pnCCD camera attached to the microscope (see Visualization 1). Thus, technically pump-probe experiments on femtosecond time scales would be feasible with the presented setup.

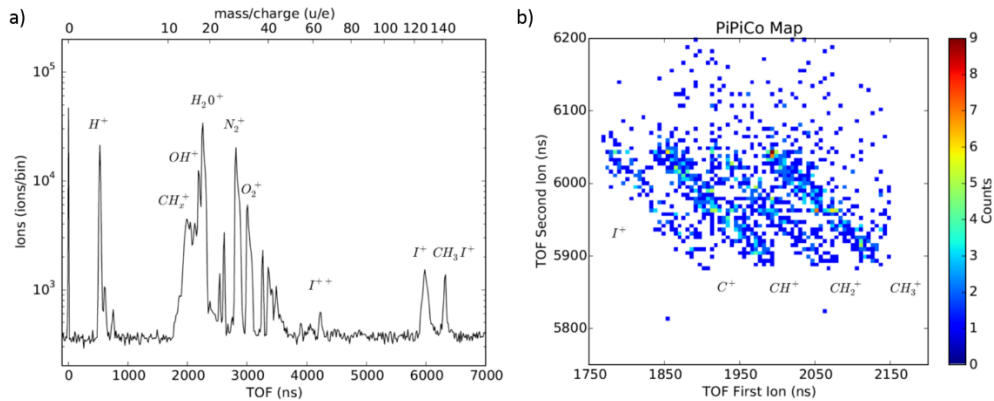

**Figure 6. a)** Ion time-of-flight mass spectrum of $CH_3I$ recorded at 68.6 eV photon energy. The $CH_3I^+$ parent ion as well as several ionic fragments can be seen along with several ions resulting from the ionization of the residual background gas in the chamber. **b)** Ion-ion coincidence spectrum of $CH_3I$ zoomed in on the region containing the coincidences between $CH_x^+$ (x = 0, …, 3) and $I^+$ fragments.

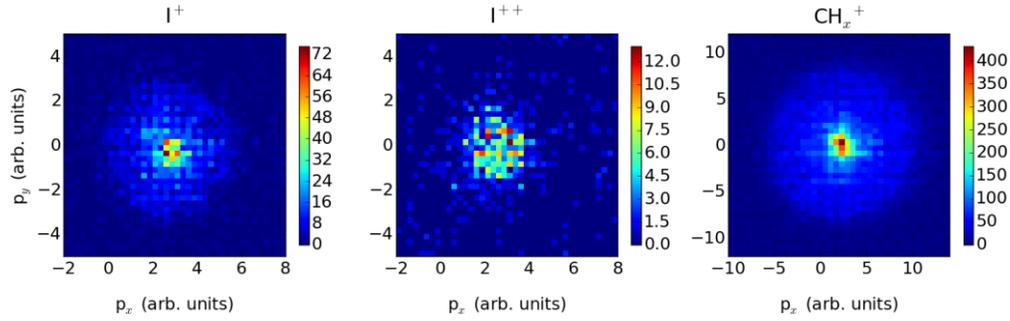

**Figure 7.** Two-dimensional ion momentum distributions (ion detector images) for three distinct ionic fragments of $CH_3I$: **a)** $I^+$, **b)** $CH_x^+$, and **c)** $CH_3I^+$.

*3.1 Photon flux estimate*

In order to obtain an experimental estimate of the actual XUV photon flux in the CAMP chamber, we leaked $5.4 \cdot 10^{-6}$ mbar of neon (as measured by a hot cathode ion gauge, taking into account the appropriate correction factor) into the vacuum chamber through a needle valve and measured the count rate of $Ne^+$ ions with our VMI spectrometer described above. Using the formula

$$I = \phi \sigma D \delta L$$

where I = 2500 Hz is the $Ne^+$ ion count rate, $\phi$ is the photon flux, $\sigma$ = 6.1 Mbarn is the Ne photoionization cross section at 69 eV [63] (which we consider to be constant over the bandwidth of the XUV pulses), D = 0.5 is the detector efficiency, $\delta = 1.3 \times 10^{11}$ cm$^{-3}$ is the target density, and L=2.0 cm the interaction length inside the spectrometer, we can calculated the photon flux to be $\phi = 3.2 \times 10^9$ photons/s.

Note that the uncertainties connected to the ion gauge reading, the exact detector efficiency, and the effective interaction length that is imaged by the spectrometer, result in an uncertainty of about a factor of two for the absolute photon flux. When taking into account the estimated losses that occur on the path into the CAMP chamber (EUV mirror reflectivity $(R=0.5)^2 = 0.25$; transmission of the aluminum filter of 0.5 and of the grazing incidence plates of 0.4; transmission through the HHG chamber due to absorption because of the high background pressure of 0.3), we estimate a photon flux of about $2 \cdot 10^{11}$ photons/s that was generated by the HHG source. This even exceeds the photon flux measured with a grating spectrometer as described in section 2.

## 4. Conclusion and Outlook

In conclusion, we demonstrated coincidence experiments on gas phase-molecules after inner-shell ionization with a table-top high harmonic source. This XUV source, based on a femtosecond fiber laser system, provides a record high photon flux ($> 4 \cdot 10^{10}$ photons/s per harmonic) up to 71 eV at up to 100 kHz repetition rate. The unprecedented combination of high photon flux, high repetition rate, and long term stability enables photon-hungry experiments such as photoionization of gas-phase molecules with coincidence detection of the fragments. We have performed molecular physics experiments with continuous experimental runs up to 20 hours. Electron-ion-ion coincidence data were recorded for the photoionization of $CH_3I$ molecules excited by HHG pulses at a photon energy of 68.6 eV. This first data is promising and indicates that pump-probe experiments with such table-top sources are in reach. Since the pump pulse and the XUV pulse are generated by the same laser and are, thus, inherently synchronized, it is technologically possible to control their relative timing with sub-fs precision as demonstrated by attoseocond pump-probe beamlines in many laboratories worldwide [64]. Note that a 600 kHz repetition rate attosecond source has already been demonstrated [49], and laser technology is currently progressing towards much higher average power for few-cycle driving lasers [38,65].

However, for real pump-probe experiments, the count rate of the presented experiment has to be increased by about two orders of magnitude. We expect to obtain at least an order of magnitude higher sample density with a more compact molecular beam design that greatly reduces the distance between nozzle and interaction region. The anticipated reduction in electron energy resolution due to space charge effects is far less critical compared to the significant improvement in count rate, which would allow for pump-probe scans with reasonable measurement times (few-hours). Furthermore, the experiments presented in this manuscript have been performed with 25 W (50 W) of average power. Since more than 500 W of average power have been demonstrated from a femtosecond fiber laser already [30] an order of magnitude higher photon flux is expected to be available soon.

Finally, the combination of both improvements will enable ultrafast time-resolved imaging of charge transfer processes with unprecedented time resolution on time scales which are hard to be accessed by todays FEL facilities.

**Acknowledgements**

We kindly thank Fabian Stutzki, Martin Gebhardt and Michael Müller for assistance on fiber preparation, Stefan Demmler for help with the pulse shaping system and Marco Kienel for support on the coherent combining. We also thank Achim Czasch (Roentdek GmbH) for providing routines and support for the COBOLD data acquisition and analysis software package.

**Funding**

This work has been partly supported by the German Federal Ministry of Education and Research (BMBF) with project 05P2015 - APPA R&D: Licht-Materie Wechselwirkung mit hochgeladenen Ionen , and project 13N12082 – NEXUS, the European Research Council under grant agreements no. 617173 - ACOPS, 240460 - PECS and 614507 - COMOTION, the excellence cluster ``The Hamburg Center for Ultrafast Imaging -- Structure, Dynamics and Control of Matter at the Atomic Scale" of the Deutsche Forschungsgemeinschaft (CUI, DFG-EXC1074), the Helmholtz Young Investigator program and the Helmholtz Virtual Institute "Dynamic Pathways in Multidimensional Landscapes". DR acknowledges support through the U.S. Department of Energy, Office of Science, Office of Basic Energy Sciences, Division of Chemical, Geological, and Biological Sciences, under contract No. DE-FG02-86ER1349.